**Value-level data provenance in an electronic laboratory notebook used by humans and language-model agents: An experience report**

Oleksiy V. Penkov[1,3] and Boyang Shen[1,2]

[1]ZJU-UIUC Institute, International Campus, Zhejiang University, Haining, 314400, China

[2]School of Mechanical Engineering, Zhejiang University, Hangzhou, 310030, China

[3]Department of Mechanical Science and Engineering, University of Illinois Urbana-Champaign, Urbana, IL, 61801, USA

Correspondence: Oleksiy V. Penkov, oleksiypenkov@intl.zju.edu.cn, ORCID 0000-0003-3675-5301

Short title: Value-level data provenance in an electronic laboratory notebook

Article category: Experience report

# Abstract

A published laboratory number can be verified only when its raw file and extraction method are known. This paper presents the value-level provenance architecture of APS ELN, an electronic laboratory notebook used by both researchers and language-model agents. The architecture was motivated by a production audit in a materials laboratory covering projects performed over almost 5 years. Tracing eight baseline studies from sample to stored value showed that the initial ELN structure preserved the link from sample to raw file. Still, the link from raw file to stored value was lost, mainly due to human factors: manual entries carried no file reference or extractor version. To close this gap, the database design and workflow were reworked so that every stored value records its source file, the data file reader/viewer (DFR) plugin and version, any interactive analysis selection, or the server-side formula for derived quantities. An automated drift check re-runs each DFR on the stored files. In two end-to-end validation studies recorded by agents in X-ray reflectivity and tribology, all stored values retained verifiable provenance with zero drift. A comparison with five other platforms shows that none binds individual stored numbers to both their source files and extractor versions.

**Keywords:** electronic laboratory notebook; data provenance; research data management; FAIR data; language-model agents; materials laboratory

**Practitioner points**

1. Hierarchical record trees in an electronic laboratory notebook preserve the link from sample to raw measurement file by structure, because files are attached under their sample and operation nodes. In contrast, numerical values entered alongside those files lose their provenance unless the schema binds each value to its source file.
2. Verifying a stored number against its raw file requires recording the source file identifier, the data file reader/viewer (DFR) plugin and version, any interactive analysis selection, or the server-side formula and input values for derived quantities, so that an automated drift check can re-evaluate the file and flag stale numbers.
3. Language-model agents write laboratory records an order of magnitude faster than users and naturally cite source files in free text; exposing structured provenance fields in the programmatic writing interface converts those text citations into queryable, verifiable links.

# 1. Introduction

A laboratory result is a chain of synthesis, a sample, an operation on it, a raw file, and a number. The number goes into a paper, while its link back to the raw file and to the method that produced it allows the result to be checked. The FAIR Guiding Principles require that research data be Findable, Accessible, Interoperable, and Reusable for both machines and people, with principle R1.2 specifically requiring that data be associated with detailed provenance [1]. Materials science has adopted these principles as a foundation for its data infrastructures [2]. Yet, an electronic laboratory notebook (ELN) can store every element of an experimental chain while still losing the links between them. Empirical studies of computational reproducibility reveal a similar failure mode when links between data, code, and execution context are lost. In a random sample of 204 papers published in *Science*, artifacts were obtained for 44% while findings were reproduced for 26% [3]. Among 10,388 biomedical Jupyter notebooks whose dependencies could be installed, only 1,203 executed without errors [4]. Similarly, only 12.0% of R scripts archived on Harvard Dataverse ran without errors, most frequently failing due to missing files, library versions, or working-directory paths [5]. For an experimental notebook, practical reproducibility likewise depends on which provenance links the system preserves by structure.

Existing ELNs and research-data platforms store both files and numerical values, but they attach provenance at different levels of granularity. In eLabFTW [6] and RSpace [7], numerical values are entered into form or record fields alongside attached files without a structural link between a field and a specific file. In openBIS [8], scripts executed at registration assign key-value properties to a dataset comprising a collection of files, whereas Chemotion ELN [9] populates record fields from instrument files through a file converter. NOMAD [10] parses uploaded raw files automatically and records, for each upload entry, the main file, the parser name, and the platform version. Because those parsers are complex software artifacts in which grammar-based fuzzing has uncovered logic defects [11], tracking the extractor version is essential when stored results are re-evaluated. Although these architectures define where metadata can reside, they do not show which links are preserved in daily practice when values are entered, copied, or revised by researchers and by autonomous software agents. Language-model agents now plan and execute chemical syntheses through tool interfaces [12,13]. In addition, autonomous laboratories have operated continuously over 17-day experimental campaigns [14], introducing high-frequency programmatic writes into databases originally designed around manual entry.

In this study, value-level data provenance and automated drift detection were designed and implemented in APS ELN [15]. Seven months of production operation in a materials laboratory were first audited. During this period, records were written both by users at desktop and web clients and by language-model agents operating through a Model Context Protocol (MCP) server [16–19]. Tracing eight baseline studies identified where unlinked records lose provenance (Section 2). Seven architectural extensions were then implemented to bind every stored value to its source file, data file reader/viewer (DFR) plugin and version, interactive analysis selection, or server-side formula (Section 3). Finally, the architecture was validated end-to-end in two agent-recorded studies in X-ray reflectivity and tribology with independent raw-file verification (Section 4) and compared against five external platforms (Section 5).

The contributions of this paper are fourfold:

1. An empirical audit of provenance preservation across eight traced baseline studies in seven months of production database use by humans and language-model agents (Section 2).

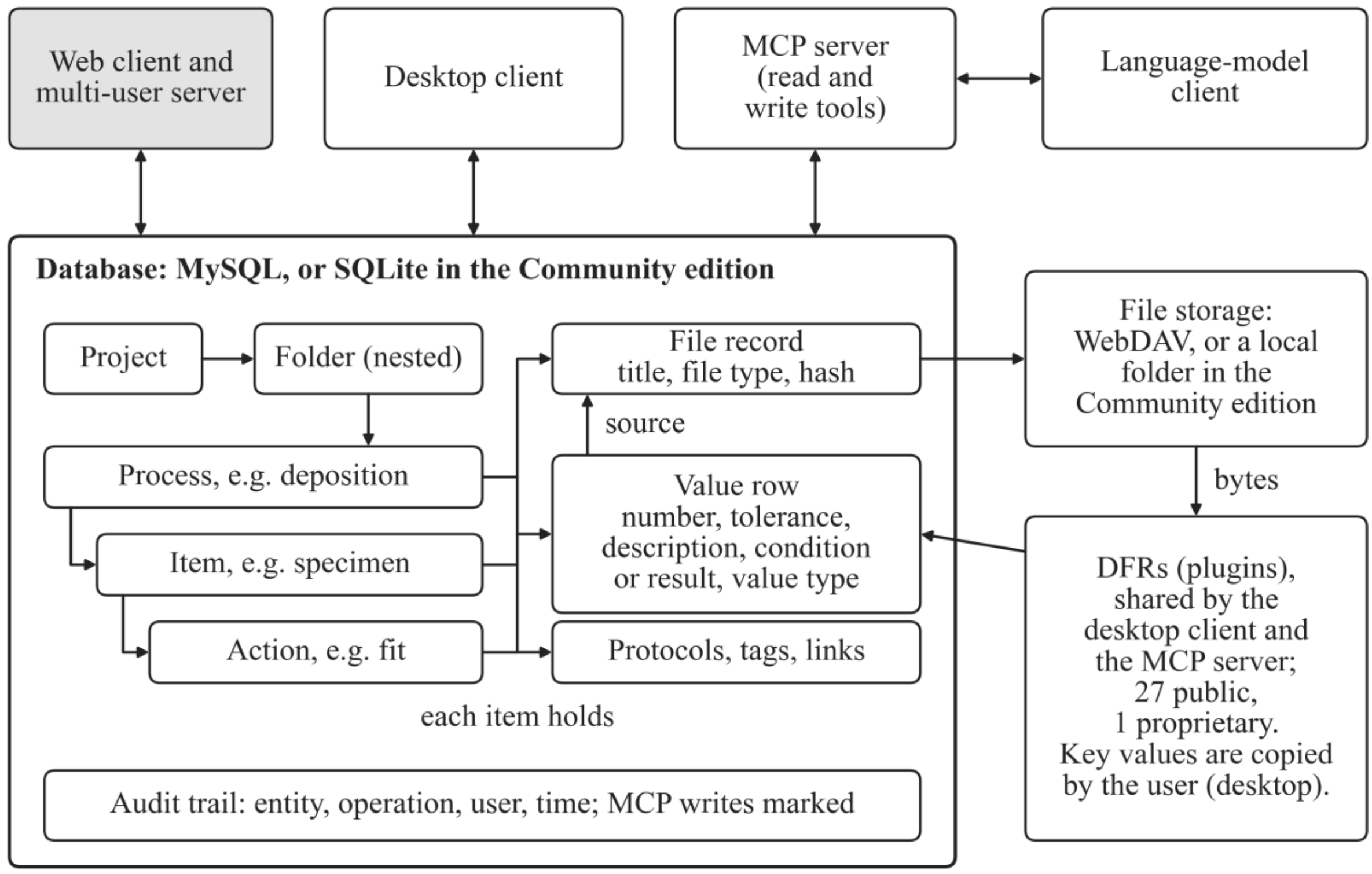

**Figure 1.** Data model and clients of APS ELN. Each node in the tree holds file records, value rows, protocols, tags, and links. The multi-user installation studied here stores records in MySQL and file bytes in WebDAV storage, whereas the Community edition stores them in a local SQLite notebook and a local folder. File bytes are read by data file readers/viewers (DFRs). The arrow labeled "source" links a value row to the file it came from (Section 3.1). Gray: not in the public edition.

2. The design and open-source implementation of value-level provenance covering reader-extracted values, interactive analysis selections, and formulas derived server-side, paired with entry validation, content hashing, and automated drift detection (Section 3).
3. End-to-end validation of the architecture through two agent-recorded studies in X-ray reflectivity and tribology, establishing complete provenance retention and zero drift and resolving the baseline provenance gaps (Section 4).
4. A comparative study of five established ELNs and research-data platforms against the same value-level provenance criteria (Section 5).

Section 2 describes the APS ELN system, its seven-month production setting, and the baseline audit that motivates the design (Sections 2.1 to 2.3). The value-level provenance architecture and evaluation protocol are presented in Section 3 (Sections 3.1 to 3.4). Section 4 reports the results of the two end-to-end validation studies and the resolution of the baseline provenance gaps (Sections 4.1 to 4.3). Related systems are compared in Section 5, while threats to validity and conclusions follow in Sections 6 and 7.

## 2. System overview and motivation

This section describes the APS ELN installation studied here and the baseline audit performed on its production database. The audit traced eight studies from sample creation to stored numerical values. Its findings motivate the value-level provenance architecture presented in Section 3.

## 2.1 Clients, data model, and DFRs

APS ELN is an electronic laboratory notebook designed for experimental materials research. In the production installation studied here, three clients (a Windows desktop application, a browser-based web client, and an MCP server) operate on a shared MySQL database while storing attached file bytes in WebDAV storage. MCP enables a language-model client to query and write notebook records by structured tool calls. The desktop client and the MCP server are released as open-source software in the Community edition [15], which stores the same data model in a local SQLite notebook and a local file directory. The web client and multi-user server remain proprietary.

As shown in Figure 1, the data model organizes each project into folders containing a hierarchical tree of typed items. Synthesis or fabrication processes sit at the root of the tree, specimens are found beneath the processes that produced them, and measurement or analysis actions are attached to the corresponding specimens. By construction, this hierarchy records which process produced a sample and which characterization operations were performed on it.

Each item record stores a title, a free-text description, and key-value properties, while owning child collections of files, numerical values, protocols, tags, and cross-tree links. A file record stores a title, a file type, an optional content hash, and a foreign key to its parent item. The file bytes reside in storage at a path constructed from the project, item, and file name. A value row stores a floating-point number, an optional tolerance, a free-text description, and a condition-or-result flag. It also references one of 36 predefined value types that specify the physical quantity and its canonical unit. In the initial schema examined in Section 2.3, a value row was linked only to its parent item and contained no field referencing a source file, an extractor plugin, or an extractor version.

Protocols attach structured equipment forms with validation states to items. Tags classify items across projects, while links connect related items through different branches of the tree. Every database write is recorded in an append-only audit trail with the entity type, entity identifier, operation, user account, and timestamp, with writes originating from the MCP server marked in the audit description. Regulated pharmaceutical audit trails must preserve replaced values to reconstruct historical states [20,21], and event-sourced systems rebuild state from the event log [22]. In contrast, the APS ELN audit trail is an operational write log: it dates every file and value and separates MCP writes from manual writes, but it does not retain overwritten numbers (Section 5).

Instrument and analysis files attached to items are rendered and parsed by compiled plugin modules called data file readers/viewers (DFRs). Each file is routed to the DFR whose declared extension yields the longest match against the file name, so compound extensions take precedence over single-part extensions. Beyond displaying the file in the desktop interface, a DFR can provide a thumbnail preview, a list of extracted key values, a structured document tree, and tabular data columns. Each extracted key value comprises 6 fields: a parameter name, a numerical value, a unit string, a description, a condition-or-result flag, and a headline flag. Because the desktop client and the MCP server dynamically load the same compiled DFR packages, both interfaces extract identical numbers from a given file.

In the desktop client, a user can import extracted key values into an item's value table by mapping each selected key to one of the 36 value types, which inserts a new value row. Through the MCP server, a client can query the key values extracted from each attached file alongside the values already stored on the item. The installation deploys 28 DFRs, of which 27 are included in the Community edition [15], and one (the proprietary sputtering-log DFR) remains internal. Consequently, any stored number derived from a supported file format can be checked by re-running the current DFR on the stored file bytes.

### 2.2 Production setting and baseline audit method

The production installation serves a single materials research laboratory, in which it replaced an earlier custom notebook (ELN2). Over seven months of operation from March to October 2026, the audit trail recorded 7710 writes across 24 visible projects containing 1283 items, 2676 files, and 674 values. This corresponds to roughly two attached files per item, but only one stored numerical value for every two items. Records were entered by seven people using the desktop and web clients and by language-model clients responsible for 2043 writes (26%) through the MCP server under the first author's account and a dedicated agent account. The agent account was shared by an autonomous experimental agent and an orchestrator process that wrote notebook records from the agent's session logs.

Programmatic and manual workflows operated at markedly different speeds. A working session was defined as a sequence of writes separated by at most 30 min. Within sessions, consecutive writes through the MCP server occurred at a median interval of 1 s under the first author's account and 2 s under the agent account. Across the six users who performed more than 100 manual writes, the median interval between consecutive writes ranged from 9 to 23 s. Thus, agent-driven recording was roughly an order of magnitude faster than manual entry.

Part of the production database was migrated from the legacy ELN2 system, which operated from February 2021 to early 2026. ELN2 accumulated 19 projects (initiated between February 2021 and July 2023) containing 1971 specimens, 1794 experiments, 2034 tests, 3558 test files, and 3092 experiment files. ELN2 introduced the laboratory's sample-naming convention and holds the experimental records underlying three published coating studies on SiNx/BN [23], diamond-like carbon (DLC)/B4C [24], and W/B4C [25]. In each case, ELN2 contains specimen series corresponding to the published multilayer period (the thickness of one repeated bilayer) and the varied deposition parameter. Those varied parameters were pulse frequency and thickness ratio for SiNx/BN, nitrogen flow for DLC/B4C, and tungsten current corresponding to the published tungsten power series for W/B4C. However, ELN2 stored no explicit link between specimen records and publications. A subset of these legacy results was imported into APS ELN through the MCP server under the first author's account in a single batch of 269 writes. That batch created 91 items comprising 85 specimens, their test actions, and one summary page per group, with sample codes dated between December 2023 and April 2025. In three APS ELN projects, the specimen descriptions reference the first ELN2 records in free text (verified against ELN2 for three sampled specimens). This import accounts for sample codes predating the audit trail and for 269 of the 2043 MCP writes.

All production data were inspected in read-only mode via the MCP server using tools that retrieved database records, audit entries, and extracted file contents without modifying the notebook. Because one project in the database was restricted from the audit account, all reported totals refer to the 24 visible projects. The complete audit trail provided the creation timestamp of every item, file, and value. It also separated manual writes at the desktop and web clients from programmatic writes through the MCP server.

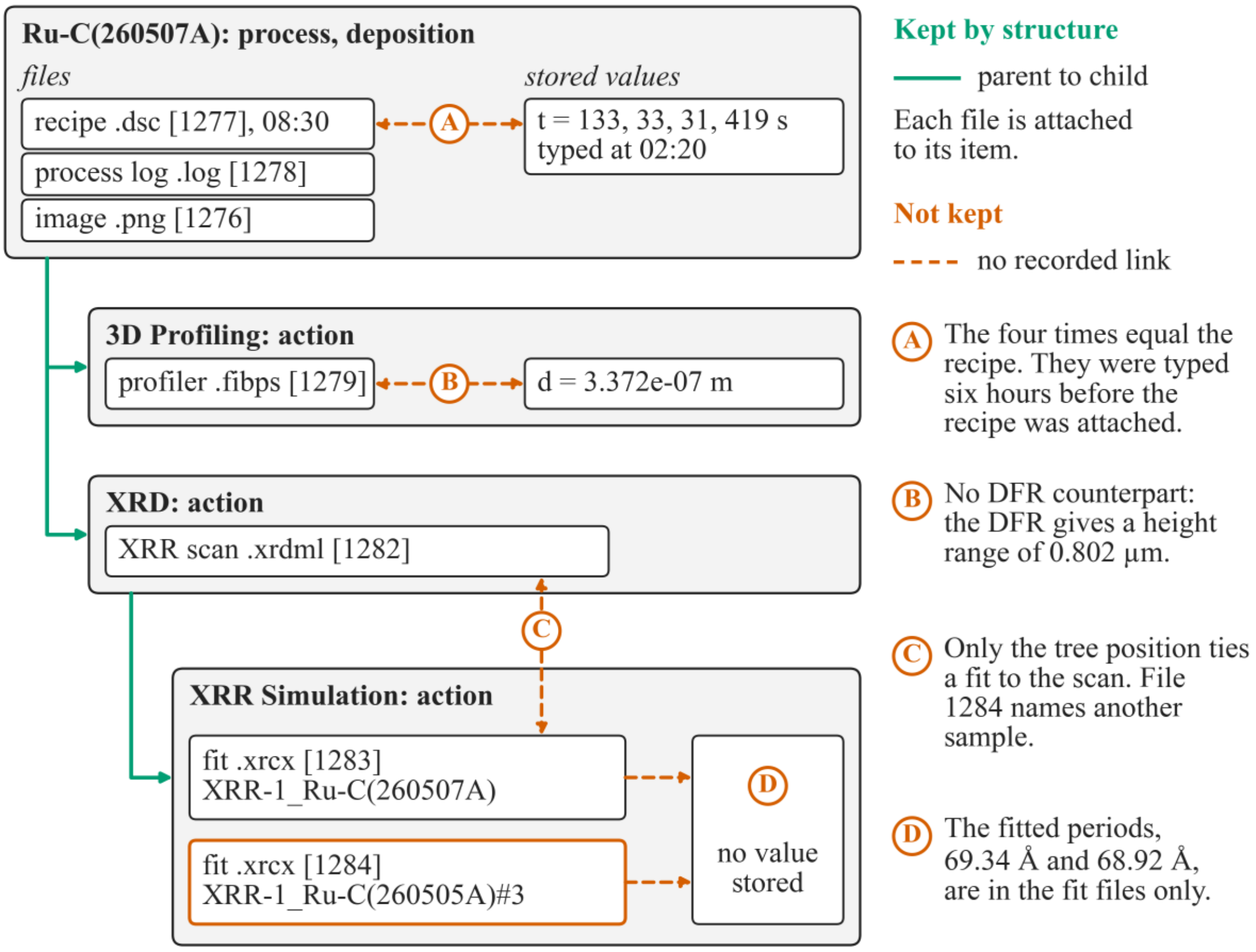

**Figure 2.** Study T6, a Ru/C multilayer, as stored. Gray boxes are items, while white boxes are their files (IDs in brackets) and stored values. Solid arrows: tree links. Dashed arrows A–D: links the record does not hold (Section 2.3). DFR, data file reader/viewer.

To determine which provenance links were preserved without value-level binding, eight experimental studies (T1–T8) were traced from sample creation through stored numerical values across five projects. Three projects representing general coatings, etching, and tribology workflows were first examined at the item level. Two were extracted in full, while the largest project (432 items) was sampled by selecting every eighth root process (15 root processes with their complete subtrees). Within each of these three projects, the study with the largest subtree containing stored numerical values was traced in full (T1–T3). Three additional handwritten studies were selected from X-ray optics projects (T4–T6), each comprising a magnetron sputtering deposition, a profilometry thickness measurement, an X-ray reflectivity (XRR) scan, and an XRR curve fit used to determine layer thicknesses. Finally, one complete deposition-measurement-fit chain was selected from each of the two projects recorded by language-model agents (T7 and T8).

For every traced study, each stored numerical value was compared against the key values extracted by the DFRs from the files attached within the same study, referred to below as the DFR output. Each stored value was classified into one of three mutually exclusive categories. A value was *equal to DFR output* when the stored number agreed with an extracted key value of an attached file after unit conversion. It was *different from DFR output* when an attached file's DFR extracted the same physical quantity but returned a different numerical value. It had *no DFR counterpart* when no DFR for any file attached in the study extracted that quantity.

Data extraction and verification were performed with the assistance of a language-model coding agent connected to the MCP server. Operating under selection rules and classification criteria defined by the first author, the agent computed the audit statistics and write-interval distributions reported in this section. It also matched stored values against DFR outputs for Table 1 and inspected the documentation and source repositories for the five comparison systems in Section 5. The first author independently verified every row of Table 1 against the notebook records and checked all system comparisons in Section 5 against the primary documentation and source code.

### 2.3 Motivation: provenance gaps without value-level binding

Across the six handwritten studies (T1–T6), the item tree preserved the link between each specimen, its processing steps, and its raw files, whereas it could not record which scan a downstream fit analyzed (Figure 2). Measurement actions were attached beneath the corresponding specimen item or, when no separate specimen item was used, beneath the root synthesis process. In the X-ray optics studies (T4–T6), the root deposition item held the recipe and sputtering log, while each XRR fit action was nested beneath the measurement action that it analyzed. Where a sample-to-file chain was incomplete, the defect consisted of missing attachments under properly placed nodes. Placeholder actions were left empty in T1, T3, and T5, while two screen captures were attached in place of the raw XRR scan file in T5. In T6, the fit action held two XRR fit files, one of which carried the code of a different specimen in its title. Neither the action nor the file records encoded which fit was accepted (links C and D in Figure 2). Thus, an unrelated or superseded fit could sit under a measurement node without detection.

The next link, from attached file to stored number, was lost across the handwritten studies (Table 1). Of the 26 numerical values stored in T1–T6, 13 equaled the DFR output of an attached file, 4 differed from it, and 9 had no DFR counterpart. Agreement varied sharply by domain, with 11 of 14 stored values matching DFR outputs in T4–T6 compared with only 2 of 12 in T1–T3. None of the 26 values had been entered through the MCP server. Because the schema provided no foreign key to a file record, audit timestamps showed that even matching values had been typed manually rather than imported from a DFR (link A in Figure 2).

Table 1. Stored values of the eight traced studies by comparison category. The category compares each value with the output of the data file readers/viewers (DFRs) for the files attached in the same study.

| Studies | Values | Equal to DFR output of an attached file | Differ from it | No DFR counterpart |
|---|---|---|---|---|
| T1–T3, coatings, etching and tribology | 12 | 2 | 4 | 6 |
| T4–T6, X-ray optics | 14 | 11 | 0 | 3 |
| Total, handwritten | 26 | 13 | 4 | 9 |
| T7, written by the agent | 14 inputs, 0 results | – | – | – |
| T8, written by the orchestrator | 0 | – | – | – |

Without value-level provenance and entry validation, duplicate, contradictory, and improperly scaled numbers also persisted in the record. On one nanoindentation action in T1, two pairs of hardness and modulus values coexisted without indicating which pair was current. The first pair had been typed before file upload and disagreed with the DFR output, whereas the second pair matched the DFR output of both attached indentation files. A coating thickness in T1 was recorded as 319.4 with the unit set to meters ("m"). Similarly, tolerances in T2 and T5 were entered on scales incompatible with their values (0.060 on $2.28 \times 10^{-5}$ m, and 10 on $1.49 \times 10^{-7}$ m). In addition, none of the profilometry film thicknesses in T1–T6 matched a DFR key value, because the profilometry DFR returned only the total height range (link B in Figure 2). In the Si-DLC study (T3), a hardness of 14.1 GPa and a modulus of 1170 GPa were entered at upload, whereas re-reading the stored file with the DFR returned 8.97 GPa and 1330 GPa. Because value rows recorded no DFR version and 62% of database files carried no content hash, the database alone could not distinguish a modified file, a revised DFR algorithm, or a different indent selection.

The two agent-recorded studies (T7 and T8) inverted this pattern. In T7, the autonomous deposition agent [16] recorded a Ru/C multilayer, while in T8 the orchestrator process recorded a Co/C multilayer. Both agents wrote the source file name, job identifier, and accepted-fit identifier for every reported number into the item text. Spot checks of the top period in T7 and of the period and two layer thicknesses in T8 matched the fit files in every case. However, both workflows bypassed the value table for experimental outcomes: T7 stored 14 input rows and 0 results, whereas T8 created 79 items and attached 397 files with 0 value rows. In addition, the 15 programmatic deposition runs in T7 lacked console snapshots, and the raw recipe and log files had no corresponding DFR to extract key values. Thus, closing these gaps requires binding each stored number to its source file, DFR version, analysis selection, or formula while exposing structured provenance writes to language-model agents (Section 3).

## 3. Value-level provenance architecture

The software [15] implemented seven extensions which jointly form a closed provenance loop around the value source record (Section 3.1). Structured MCP provenance writes (Section 3.2) and DFR plugins for deposition control-system formats (Section 3.3) populate this record at entry, while analysis snapshots and server-side derived formulas (Section 3.3) extend it to interactive selections and multi-file calculations. Entry validation and content hashing (Section 3.2) close the unit-scale and storage-integrity gaps of Section 2.3. In contrast, the automated drift check (Section 3.1) replays each stored extraction, selection, or formula against the raw files.

### 3.1 Value source records and automated drift detection

To resolve the missing file links and unexplained value discrepancies of Section 2.3, the software attaches a structured source record to every value row [15]. This record expresses the derivation process of each value from its primary source [26]. Each value has a source kind of "Reader", "Derived" (Section 3.3), "Manual", or "Unknown". A "Reader" source denotes a value extracted by a DFR and holds four fields: source file ID, DFR plugin ID, DFR version, and extracted key name. A value entered directly by a user is marked "Manual", while editing a "Reader" value changes its source kind to "Manual" and retains a note of the earliest DFR. Rows written without a source record are reported as "Unknown" because their provenance cannot be reconstructed retroactively.

DFR plugins are research software whose behavior can change between releases. The FAIR principles for research software require that distinct versions carry distinct identifiers [27]. Each DFR exposes a plugin identifier and a version, which 20 of the 27 public DFRs report. The remaining DFRs report an unknown version until they are stamped. An automated, read-only drift check re-runs the corresponding DFR on the stored file bytes of every "Reader" value. The check flags a row when the re-extracted value, the DFR identifier, or the DFR version has changed. It also flags a row when the key is no longer returned or when the source file is missing or unreadable.

### 3.2 MCP provenance writes, entry validation, and content hashing

The write tools of the MCP server accept the structured source object of Section 3.1 alongside each value. This converts the free-text file citations that agents wrote in Section 2.3 into queryable database links. The referenced source file must belong to the target item, to one of its ancestors, or to one of its descendants in the same project, which preserves hierarchical locality in the project tree. A value written without a source object defaults to "Unknown". A batch write tool commits up to 200 values in one atomic transaction, so all parameters of an analysis or a curve fit enter the database together. The tool descriptions instruct agents to write fit and analysis outputs as source-linked value rows rather than only in item descriptions.

Administrators can define optional minimum and maximum plausible bounds in canonical units for each value type. A relative tolerance check flags any entry whose tolerance exceeds |value| or falls below |value|/1000, with a scaling factor that is configurable per database. The check tests the order of magnitude of the tolerance without converting units. When an entry violates either rule, the desktop client asks the user for explicit confirmation. Automatic batch extraction skips the value, and the MCP server rejects the write unless the caller sets an explicit override. A read-only audit tool scans existing records for violations. The erroneous 319.4 "m" coating thickness of Section 2.3 carried a tolerance of 15.3, which lies between |value|/1000 and |value|. Thus, the tolerance rule alone cannot flag a unit-selection error, and the value type needs plausible range bounds.

The software computes and records an MD5 content hash at file upload. A read-only storage content check re-reads the physical file bytes and classifies each file record as intact, missing from storage, mismatched against its recorded hash, or unhashed. The check can also re-evaluate a list of previously flagged files to track remediation. For unhashed files in the production database (Section 2.3), the check confirms physical presence, whereas for hashed files it verifies byte-level integrity.

### 3.3 Control-system DFRs, analysis selections, and computed values

The software implements three DFR plugins for the deposition control-system formats of Section 2.3 [15]. The recipe DFR parses native recipe files and returns every numeric set point as an experimental condition in the declared unit of the file. It does not simulate the script to infer per-layer thicknesses or loop counts. The event-log DFR extracts the total run duration, the counts of safety-interlock trips and incomplete steps, and the final run outcome. The data-series DFR exposes time-series sensor columns for plotting. When paired with a run snapshot, it returns 48 key values, including power, time, current, pressure, and repeat-count set points, for a complete deposition snapshot. A time-series table without a snapshot yields plotting columns only. Recipes saved as plain text fall back to the generic tabular DFR. Thus, automated verification of deposition set points requires archiving either the run snapshot or the native recipe file.

Several quantities depend on an interactive selection in addition to the raw file and the DFR version. The step height from optical profilometry depends on two reference rectangles chosen on a leveled surface map. Nanoindentation hardness and modulus depend on which individual indents are included. The friction coefficient of a reciprocating test depends on the cycle window averaged after running-in. The software stores such choices on the value row as a per-value JSON analysis snapshot [15]. A snapshot has eight base fields: schema version, DFR identifier, DFR version, algorithm version, file identifier, file MD5 hash, key name, and selection mode. The selection mode is "automatic" when the DFR determined the region without user input. Domain coordinates complete the snapshot: two pixel rectangles and the map dimensions for the step height, the active indent indices for nanoindentation, or the cycle interval for friction. During the drift check, the server passes the stored snapshot back to the DFR. The DFR then replays the extraction with the exact selection that produced the value.

For a value with a snapshot, the drift check reports one of four statuses. An "unchanged" status means that the replayed extraction reproduces the stored value, whereas "selection changed" indicates that the default selection currently saved for the file differs from the stored snapshot. A "source changed" status means that the current MD5 hash of the file differs from the hash in the snapshot. Finally, "unusable" means that the stored coordinates cannot be applied to the file (for example, an indent index out of range or a rectangle outside the height map). A value stored without a snapshot falls back to the current default selection of the file. Thus, the snapshot status separates a modified file from an altered selection and from a changed DFR.

To enable autonomous workflows to record selection-dependent quantities without GUI interaction, the MCP extraction tool accepts an optional per-call selection parameter. The DFR evaluates the specified coordinates for that call only without altering the saved desktop state of the file. It then returns the resulting snapshot so that the agent can pass it directly to the value-write tool (calls to DFRs without selection support are rejected). For optical profilometry, a companion MCP tool returns a downsampled, plane-leveled height map from which an agent can choose substrate and coating rectangles before extracting the step height.

Quantities like the specific wear rate of a tribological test are not stored in any single instrument file. Instead, they are calculated from the wear-track cross-sectional area, normal load, and number of sliding cycles across several actions. The software implemented the "Derived" source kind for such quantities [15]. Rather than accepting a client-computed number, the server evaluates a registered, versioned formula over existing value rows in the same project. Those input rows are taken from the target item or its descendants and match the physical quantity and unit required for each formula role. The resulting row records the formula identifier, the formula version, the identifiers of all input value rows, and a snapshot of the number that each input held at computation time, with no tolerance field.

Version 1 of the reciprocating wear-rate formula is $K = A / (2 F N)$, reported in $mm^3/(N \cdot mm)$. Because wear volume equals the mean track cross-sectional area A multiplied by stroke length while total sliding distance equals 2 N multiplied by stroke length, the stroke length cancels analytically. This leaves three input roles: the cross-sectional wear area A (which accepts an array of stored area values from replicate profile scans and uses their mean), the normal force F, and the cycle count N. When the drift check inspects a Derived value, it first verifies every "Reader" or nested "Derived" input against its source file. It then re-evaluates the saved formula version on the current input numbers and flags a changed output, a changed input value, or a missing input row. If the saved formula version is no longer registered on the server, the check reports that status without recomputing. An input row with a "Manual" or "Unknown" source marks that derived value as not fully re-executable from raw files, while editing a Derived value changes its source kind to "Manual".

### 3.4 End-to-end evaluation protocol

The complete architecture was evaluated in two fresh Community edition notebooks [15]: a Co/C multilayer XRR study (sample 261014B) and a BN/DLC coating tribology study (sample 261015A). Language-model agents connected through the MCP server recorded both studies from written instructions using raw instrument files staged in an upload directory. A recording agent and a dedicated fitting agent handled the XRR study, whereas one recording agent handled the tribology study. Every stored value had to come from a DFR extraction or from a server-side formula rather than from a raw number typed by an agent.

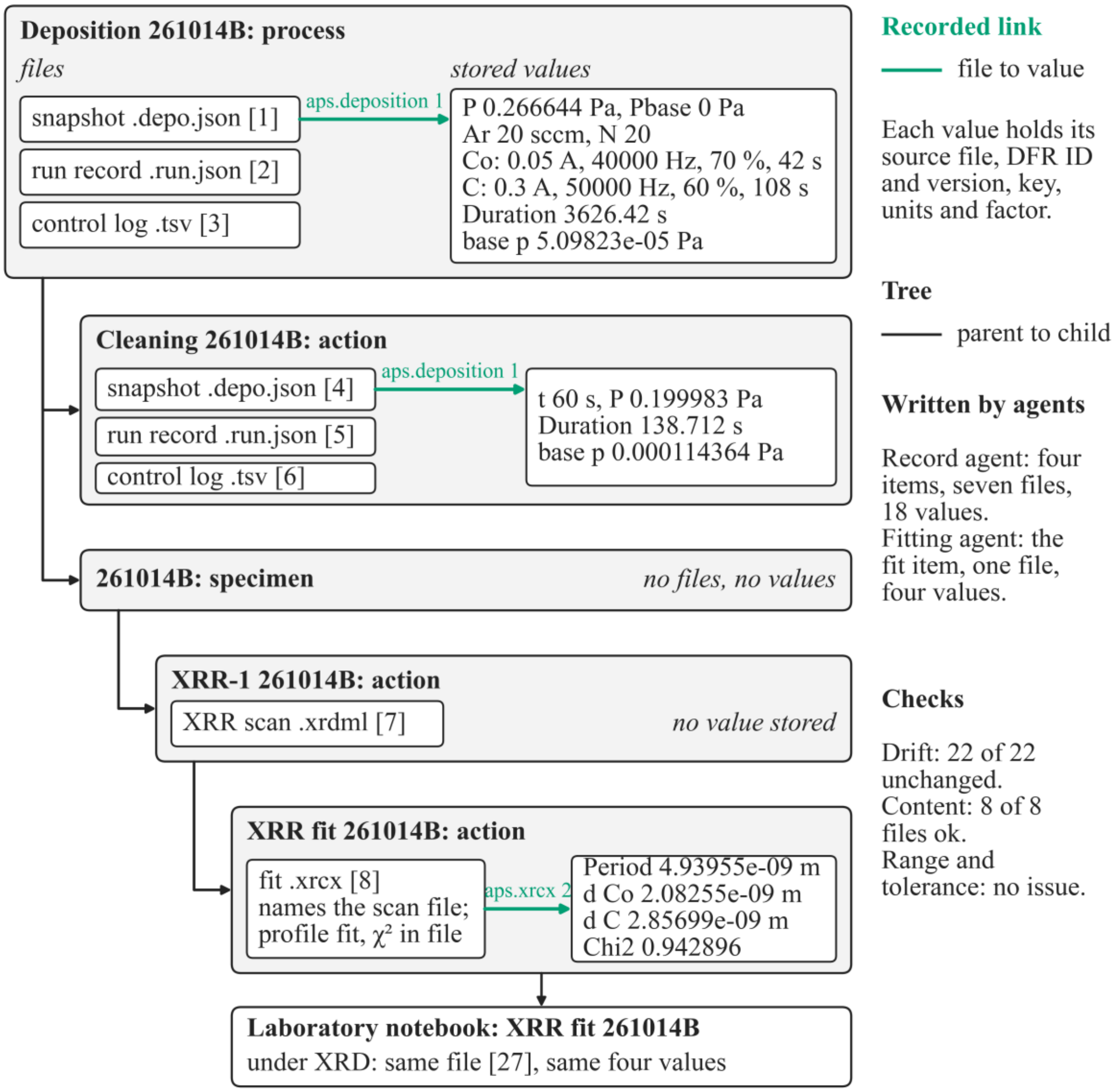

**Figure 3.** The XRR study of a Co/C multilayer, as recorded by two agents through the MCP server of the Community edition. Gray boxes are items, while white boxes are their files (IDs in brackets) and stored values. Black arrows: links in the tree. Green arrows: the source for each value row, labeled with the DFR and its version. The fit file names the scan file; the record holds no link between the two files. The laboratory notebook holds the accepted fit with the same file content and values. DFR, data file reader/viewer.

Each completed notebook was then checked with the built-in drift check, the storage content check, and the range and tolerance audit. Finally, an independent verification script confirmed file integrity by MD5 checksum and re-derived the stored numbers directly from the raw files without calling any DFR. The results of both studies are reported in Section 4.

## 4. Results

Across the two end-to-end validation studies recorded in fresh Community edition notebooks [15], language-model agents created 11 items and attached 19 files. They stored 46 numerical values comprising 45 "Reader" values and 1 "Derived" value, all of which passed the automated drift check with zero drift while every file passed the MD5 content check with zero range or tolerance violations. Independent re-derivation directly from the raw files without calling the DFRs confirmed 44 of those values to high precision and the 2 segmentation-dependent values to within an order of magnitude. Thus, every stored number retained a verifiable link to its raw file.

**4.1 End-to-end provenance in the XRR multilayer study**

In the Co/C multilayer study on sample 261014B (Figure 3), a recording agent and an XRR fitting agent created 5 items and attached 8 files, storing 22 "Reader" values. The recording agent documented substrate cleaning, magnetron sputtering deposition, and the XRR scan across 4 items, 7 files, and 18 condition and scan values. Next, the fitting agent created 1 fit action and attached 1 fit file. It recorded 4 fitted result values: a multilayer period of $4.94 \times 10^{-9}$ m, a Co layer thickness of $2.08 \times 10^{-9}$ m, a C layer thickness of $2.86 \times 10^{-9}$ m, and a goodness-of-fit $\chi^2$ of 0.94. Thus, both experimental conditions and fit results carried explicit file provenance.

Built-in verification on the completed XRR notebook confirmed all stored values as unchanged under the drift check and all files as intact by MD5 under the content check, with no range or tolerance violations. The accepted fit file and its 4 fit values written to the main laboratory notebook likewise passed without drift. An independent verification script then checked all 9 stored files across both notebooks directly against the original raw files without calling the DFRs, confirming every file by MD5 checksum and reproducing all 26 stored numbers. The 4 XRR fit parameters agreed within a relative difference of $1.3 \times 10^{-15}$, whereas the deposition parameters agreed within $2.8 \times 10^{-6}$ owing to 6-digit rounding of the pressure-conversion factors in the DFR. Thus, every number in the XRR study was independently reproducible.

**4.2 Analysis selections and derived values in the tribology study**

In the BN/DLC coating tribology study on sample 261015A (Figure 4), one recording agent created 6 items and attached 11 files, storing 23 "Reader" values plus 1 "Derived" value. Three coating step-height values were extracted from three optical-profilometry (FIBPS) files: one in automatic mode (355 nm) and two (358 nm and 316 nm) using the desktop tool's standard reference rectangles. Those rectangles spanned pixel coordinates 69 to 208 and 347 to 486 in x and 80 to 403 in y on a 556 × 484 map. They were saved as snapshots because alternative rectangle placements yielded step heights from 230 to 770 nm. Nanoindentation hardness (5.92 ± 0.47 GPa) and elastic modulus (127 ± 11 GPa) were extracted across all 9 indents with their indent-selection snapshot. The reciprocating friction coefficient (0.528 ± 0.019) was stored with its cycle-window snapshot. Finally, the server-side wear-rate formula combined the three wear-track areas (6.20, 5.53, and $6.21 \times 10^{-3}$ mm$^2$) with the 9.81 N load and 1000 cycles to record a derived wear rate of $3.05 \times 10^{-7}$ mm$^3$/(N·mm). Thus, every selection and multi-file calculation was stored on the value row.

Built-in verification in the tribology notebook reported all 24 values as unchanged under the drift check, achieved by re-executing the saved region, indent, and cycle-window snapshots and recomputing the derived wear rate, while all files passed the MD5 content check with no range or tolerance issues. Independent re-derivation of the raw files confirmed every file by MD5 and reproduced 22 of the stored values to high precision. Specifically, the 12 deposition parameters, the normal force, the cycle count, and the derived wear rate agreed within $2.8 \times 10^{-6}$. Both selection-bound step heights agreed to 12 significant digits on the stored pixel rectangles. Nanoindentation hardness and modulus matched the exact arithmetic means of the 9 per-indent values in the binary file. At the same time, the three wear-track areas agreed within 0.02–0.23% of an independent numerical integration. The remaining 2 values (the friction coefficient and the automatic step height) agreed within 0.5–4% and 10%, respectively, under simplified segmentation heuristics. Thus, saved selection snapshots and derived formulas remained reproducible from the raw files.

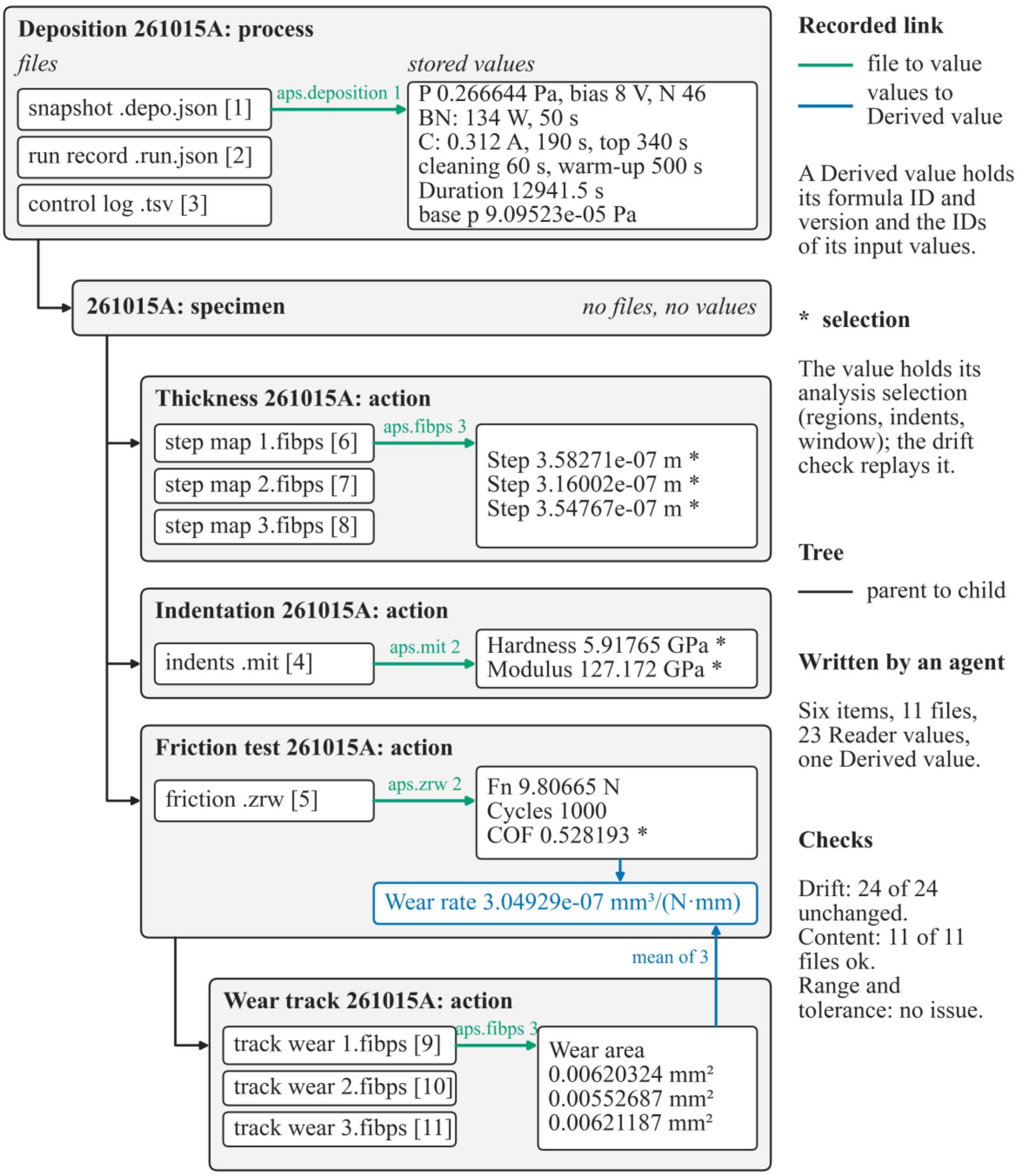

**Figure 4.** The Tribo study, a BN/DLC coating, as recorded by one agent through the MCP server of the Community edition. Gray boxes are items, while white boxes are their files (IDs in brackets) and stored values. Black arrows: links of the tree. Green arrows: the source that each value row holds, labeled with the DFR and its version. Blue arrows: the derived value, the wear rate, which holds its formula ID and version and the IDs of its three wear-area inputs, the normal force and the number of cycles. Asterisk: the value holds its analysis selection (regions, indents, or window), which the drift check replays. DFR, data file reader/viewer.

### 4.3 Resolution of the baseline provenance gaps

Across both validation studies, binding each "Reader" value to its file identifier and DFR version eliminated the unlinked manual entries and the unexplained value divergence observed in the baseline database (Section 2.3). At the same time, the source-linked MCP write tools and control-system DFRs converted the free-text agent logs of T7 and T8 into structured, queryable value rows. Storing per-value analysis snapshots resolved the profilometry region ambiguity and the nanoindentation subset ambiguity of T3. Finally, server-side formula evaluation linked the derived wear rate to its input rows across separate actions, while mandatory MD5 hashing verified every uploaded file. Thus, the architecture closed every provenance gap identified in production operation.

## 5. Related systems

Prior surveys of electronic laboratory notebooks in academic settings have established appraisal frameworks for system selection and have situated ELNs within broader research-data management workflows [28]. User studies have identified licensing costs, usability, and cross-device accessibility as primary barriers to adoption [29]. Because such reviews compare systems at the application-feature level, they do not examine whether an individual numerical value retains a verifiable link to its raw file and software extractor. To evaluate whether the provenance gap observed in Section 2.3 is specific to APS ELN or reflects a broader architectural pattern, five established ELNs and research-data platforms were examined at the level of individual stored numbers.

Each system was evaluated for two properties. The first was whether a stored numerical value records a link to its source file and to the version of the extractor (analogous to a DFR in Section 2.1) that produced it. The second was whether the system detects stale numbers when an extractor changes. The five evaluated platforms were eLabFTW 6.0.6 [6], openBIS 20.10.13.1 [8], RSpace 2.27.0 [7], NOMAD 1.4.3 [10] and Chemotion ELN 3.1.3 [9]. They were inspected by their official documentation and public source repositories at the time of submission. Table 2 summarizes the results alongside APS ELN.

Table 2. Comparison of six systems for the link between a stored number and its source file. Entries marked "not found" were searched for in the documentation and the source code of the version named. A dash marks an entry that the sources do not state.

| System | Link from a number to its file | Extractor version recorded | Stale numbers detected after an extractor change | Re-check of stored file content | Previous numbers kept |
|---|---|---|---|---|---|
| eLabFTW 6.0.6 | no | not found | – | on demand | new value only |
| openBIS 20.10.13.1 | per dataset | not found | not found | consistency check | property changes |
| RSpace 2.27.0 | no | not found | – | not found | revision per save |
| NOMAD 1.4.3 | per entry | core version only | per core version | not found | none |
| Chemotion ELN 3.1.3 | no | not found | not found | not found | full revisions |
| APS ELN 2.2 | per value | per value | drift check | on demand | no |

As shown in Table 2, none of the five comparison systems binds an individual stored number to both its source file and the version of the extractor that generated it. Four of the five platforms provide no file-level link for individual numbers. In eLabFTW and RSpace, numerical values are stored in custom fields (with units in eLabFTW and numeric bounds in RSpace) that sit alongside attached files on the same record without referencing a specific file. Neither platform includes built-in instrument-file extractors. In openBIS, key-value properties are attached to a dataset (a container that may hold multiple files) and populated at registration time by ingestion scripts (dropboxes). The property records store neither the specific source file within the dataset nor the ingestion-script identifier and version. Chemotion ELN extracts numerical values from instrument files through a file converter and a spectra service. The resulting record field stores only the extracted value and instrument identifier. Converter versions are treated as a fixed deployment property rather than per-value metadata.

NOMAD comes closest to value-level provenance. Each upload entry records the main raw file, the parser name, and the NOMAD core version that processed it. The platform can also query entries processed under older core versions for re-parsing. Even in NOMAD, however, provenance is recorded at the entry level rather than per extracted quantity. The entry metadata records the core platform version instead of an independent version for the specific file parser. Thus, a parser defect fixed without a core version bump is not reflected in the entry's provenance record. To summarize, while this comparison is limited to the five examined platforms, the absence of per-value binding to file and extractor versions is shared across all five architectures.

The six systems also make different design trade-offs regarding storage-integrity verification and value revision history (Table 2). Computing file checksums at upload is common. Chemotion hashes every file, NOMAD hashes every upload entry, and RSpace hashes inventory images. Only eLabFTW, openBIS, and APS ELN (Section 3.2) provide a built-in mechanism to re-read stored file bytes from disk and verify them against recorded hashes. Chemotion preserves a complete dataset revision on every save. OpenBIS logs historical changes for each dataset property. In contrast, eLabFTW logs only the new value of a modified custom field, while APS ELN overwrites value rows in place without retaining prior values. Retaining property revision history, as in Chemotion and openBIS, would directly clarify how duplicate or superseded entries, such as the two nanoindentation pairs in Section 2.3, arose.

More broadly, provenance architectures vary in their target granularity, representation, and capture mechanism [30]. In computational workflow environments, provenance is typically captured at the level of workflow execution graphs. For example, NaaVRE extends Jupyter notebooks into a collaborative virtual research environment that records workflow provenance using the W3C PROV Ontology (PROV-O) [31]. Scientific workflow engines such as Pegasus likewise capture execution traces but rarely record the human decision rationale behind configuration choices in structured form [32]. Designating one curve fit among several candidates as the accepted fit of record (Section 2.3) is an analogous scientific configuration decision. In the production audit, only the unstructured text of agent study T8 recorded that rationale, while inserting a structured field for the analysis of record remains an open design extension for APS ELN.

## 6. Threats to validity

Three validity dimensions limit the conclusions of this study. The internal validity of the seven-month production audit depends on how active development, deployment, and the first author's role shaped the records. Construct validity for the DFR-based and cross-platform comparisons depends on what a parser's output and a documentation search can establish. Finally, the external validity of the two end-to-end demonstration studies is bounded by their single-laboratory setting and short duration.

### 6.1 Internal validity

Because the production database examined in Section 2.3 also served as the active development deployment for APS ELN, some record anomalies (such as legacy files without hashes) may reflect transitional software states rather than user behavior alone. The audit data does not distinguish between these causes. In addition, the first author designed APS ELN and uses it in the laboratory studied, as declared in the Conflict-of-Interest statement. To reduce selection bias during the audit, the sampling and study-selection rules were fixed before inspecting the item records (Section 2.2).

### 6.2 Construct validity

Comparing stored database values with DFR outputs (Section 2.3) tests the internal consistency between a record and its attached file. It does not test the accuracy of the underlying measurement. Because DFR plugins are software parsers, a DFR can itself contain calculation defects, as demonstrated by grammar-based fuzzing of NOMAD parsers, where input-sanitization flaws caused undetected parsing errors [11]. To address parser-level construct validity in the two validation studies, an independent verification script re-derived the stored numbers directly from the raw files without using the DFRs (Section 4). Finally, the comparative study of the five external systems in Section 5 relied on official documentation and public source code rather than live test deployments, so any capability present in an uninspected module or deployment configuration is recorded in Table 2 as "not found".

### 6.3 External validity

The seven architectural extensions of Section 3 were validated end to end in two agent-recorded demonstration studies (Section 4). They have not yet been evaluated over a multi-month period of routine human and agent use. Hence, the frequency with which the drift check flags changed values during longitudinal operation remains to be measured. Within the demonstration studies, deposition conditions were extracted from run snapshots using the data-series DFR (Section 3.3). The standalone recipe and event-log DFRs were not exercised in the demonstration runs. The tribology study's reciprocating wear-rate calculation assumes a uniform track cross-section along the stroke length, which was not independently profiled outside the three measured regions. Because the empirical audit reflects seven months of work in a single materials laboratory, the quantitative rates of manual errors and agent write speeds should be tested across additional laboratories and experimental domains.

## 7. Conclusion

In this study, value-level data provenance and automated drift detection were designed and implemented in APS ELN [15]. The design was motivated by a production audit of historical data from a materials science laboratory. To close gaps identified during the audit, the database structure and workflow were reworked so that each value row was linked to a structured source record that captured the source file identifier, the DFR plugin identifier and version, any per-value analysis-selection snapshot, or the server-side formula and input values for derived quantities. Together with entry validation, MD5 content checks, and automated drift detection, this source record established a complete, machine-checkable chain from raw instrument files to stored numbers. In two end-to-end, agent-recorded studies in X-ray reflectivity and tribology, all values retained verifiable provenance with zero drift and agreed with independent raw-file re-derivation. This functionality provides per-value file and extractor-version binding that is absent in the five external platforms compared (Section 5). Evaluating its performance during extended multi-user operation is the next step for future work.

## Acknowledgements
The agent, the test sessions, and the drafting assistants are large language models (Claude Opus 5.5, Anthropic). Agents were used to collect and draft fact cards, verify numbers and cross-references against the data records, format references, and assemble the manuscript file. The agent also wrote the Python scripts that generated Figures 1 to 4. The authors reviewed and revised all AI-assisted texts and took full responsibility for the accuracy and integrity of the work.

## Funding
This research received no specific grant from any funding agency in the public, commercial, or not-for-profit sectors.

## Conflict of interest
O. V. Penkov founded and owns APS Technology, the company that develops APS ELN and publishes its Community edition. B. Shen declares no conflict of interest.

## Data availability statement
The data and scripts are archived at Zenodo (https://doi.org/10.5281/zenodo.23467070). The deposit includes a pseudonymized copy of the audit trail; the item, file, and value records for the traced projects without titles or descriptions; the per-value comparison in Table 1; and the scripts that compute the counts for Sections 2.2, 2.3, and 3.2. It also includes both validation studies for Section 4: the raw files, the complete notebooks, the agent instructions, the call logs and check reports, and the laboratory-notebook item with the XRR fit. The full production database is not public because it contains personal account data and unpublished results from the laboratory.

## Figure legends
Figure 1. Data model and clients of APS ELN. Each node in the tree holds file records, value rows, protocols, tags, and links. The multi-user installation studied here stores records in MySQL and file bytes in WebDAV storage, whereas the Community edition stores them in a local SQLite notebook and a local folder. File bytes are read by data file readers/viewers (DFRs). The arrow labeled "source" links a value row to the file it came from (Section 3.1). Gray: not in the public edition.

Figure 2. Study T6, a Ru/C multilayer, as stored. Gray boxes are items, while white boxes are their files (IDs in brackets) and stored values. Solid arrows: tree links. Dashed arrows A–D: links the record does not hold (Section 2.3). DFR, data file reader/viewer.

Figure 3. The XRR study of a Co/C multilayer, as recorded by two agents through the MCP server of the Community edition. Gray boxes are items, while white boxes are their files (IDs in brackets) and stored values. Black arrows: links in the tree. Green arrows: the source for each value row, labeled with the DFR and its version. The fit file names the scan file; the record holds no link between the two files. The laboratory notebook holds the accepted fit with the same file content and values. DFR, data file reader/viewer.

Figure 4. The Tribo study, a BN/DLC coating, as recorded by one agent through the MCP server of the Community edition. Gray boxes are items, while white boxes are their files (IDs in brackets) and stored values. Black arrows: links of the tree. Green arrows: the source that each value row holds, labeled with the DFR and its version. Blue arrows: the derived value, the wear rate, which holds its formula ID and version and the IDs of its three wear-area inputs, the normal force and the number of cycles. Asterisk: the value holds its analysis selection (regions, indents, or window), which the drift check replays. DFR, data file reader/viewer.